# Lightweight and Data-Efficient MultivariateTime Series Forecasting using Residual-Stacked Gaussian (RS-GLinear) Architecture

Abukar Ali

Queen Marry University of
London (QMUL)
School of Electrical Engineering
& Computer Science

*Abstract*—Following the success of Transformer architectures and their self-attention mechanism in language modelling—particularly due to their ability to capture long-range dependencies—many researchers have explored how these architectures can be adopted for time-series forecasting. Variants of Transformer-based models have been proposed to handle both short-and long-term sequence modeling, aiming to predict future time-dependent values from historical observations using varying input window sizes. However, despite the popularity of leveraging Transformer architecture to extract temporal relationships from set of continuous datapoints, their performance in time-series forecasting has shown mixed results. Several researchers, including Zeng et al. (2022) and Rizvi et al. (2025), have challenged the reliability of emerging Transformer-based solutions for long-term forecasting tasks. In this research, our first objective is to evaluate the Gaussian-based Linear (GLinear) architecture proposed by Rizvi et al. (2025) and to develop an enhanced version of it—referred to in this study as Residual Stacked GLinear (RS-GLinear) model. The second objective is to assess the broader applicability of the RS-GLinear model by extending its use to additional domain—financial time series and epidemiological data—which were not explored in the baseline model proposed by Rizvi et al. (2025). Most time-series implementations (Transformer-based and Linear models) we came across commonly adopt baseline codebases provided by the Hugging Face repository, including our baseline GLinear model used in this study. Therefore, the RS-GLinear model developed in this study is an extended version of the codebase introduced in the research by Rizvi et al. (2025). All Python implementations of Transformer-based models, Linear models, RS-GLinear code—including all dependencies—are available through the GitHub[1] repository links provided below.

## I. Introduction

Time series forecasting has been an important research area in many domains such as finance/economics, retail, healthcare, cloud infrastructure, meteorology, and traffic management (Toner et al. 2024). Since the introduction of Transformer Model (Vaswani et al. 2017), there has been large amount of research focusing on time-series forecasting using Large Language Models (LLM) to leverage LLM's sequential dependencies in text generation (Tan et al. 2024). Tan et al. (2024) showed that despite the recent popularity of LLMs in time series forecasting, these models (LLM) do not provide meaningful contribution to improve forecasting performance. Zeng et al. (2022) used a simple one-layer linear architecture, called LTSF-Linear, as base model to evaluate the performance of more sophisticated transformer-based long-term time series forecasting (LTSF) models. They concluded that LSTF-Linear outperformed existing sophisticated Transformer-based LTSF models in all the data sets used. Rizvi et al. (2025) performed a comparison with state-of-the-art linear architectures (such as NLinear, DLinear, and RLinear) and transformer-based time series predictor (Autoformer). Their results show that their proposed GLinear architecture, despite being parametrically efficient, significantly outperforms the existing architectures in most cases of multivariate time series forecasting. Our research has two objectives. First the project aims to develop and evaluate an enhanced version of the GLinear architecture, which we will refer to as **Residual-Stacked GLinear** model. This new variant of the GLinear architecture is designed to improve multivariate time-series forecasting by introducing deeper neural network layers and residual connections while maintaining the core features inherent to the original GLinear architecture. The modified architecture incorporates four stacked linear transformation blocks, each followed by Gaussian Error Linear Unit (GeLU) activations and dropout layers. Additionally, residual skip connections are introduced while retaining the normalization layers from the original GLinear model (section IV provides more architecture details). Therefore, the project aims to investigate whether the RS-GLinear model produces consistent improvements over the original GLinear architecture, particularly on

---

[1]*All resources needed to reproduce this work are available:*
*https://tinyurl.com/RS-GLinear, https://tinyurl.com/LSTF-Linear*
*https://github.com/t-rizvi/GLinear/*
*https://huggingface.co/docs/transformers/en/model_doc/time_series_transformer,*

datasets where the baseline model demonstrated weaker performance – most notably the ETTh1 and Weather datasets. These cases highlight the limitations of the shallow GLinear structure in capturing the complexity of temporal dependencies. Our second aim is to explore the general applicability of RS-GLinear model to two other domains that were not considered in the GLinear baseline model introduced by Rizvi et al. (2025). These additional domains we consider are:

**a)** financial time-series which are usually known to be volatile and stochastic. They tend to exhibit jumps periodically, regime changes and are non-stationary. The volatilities observed in global financial markets are influenced by factors such as macroeconomic events, geopolitical shifts, market sentiment, and unforeseeable shocks which causes periodic systemic risk on global scale.
**b)** medical time series data such as Influenza-like Illness (ILI) which are compiled by the U.S. centre of disease control and prevention. Section V provides details of all the dataset considered in this research.

## II. RELATED WORK

Most recent work involving time-series prediction on datasets exclusively from the financial market were carried out by Mozefarri, Zhang (2024). They have used mainly historical stock market indices from Yahoo Finance open-source database. The objective of their work focused on time series prediction using deep learning frameworks such as Long Short-Term Memory (LSTM) and a model based on Transformer Architecture fine-tuned for time series forecasting. They have used the encoder-decoder architecture and the multi-head attention to capture financial markets dynamics. They have also used third model such as Facebook's Prophet libraries for time series prediction due its abilities to handle non-linear trends, seasonality and non-trading days. They concluded that the Transformer-based model outperformed both LSTM and Prophet models across all stock indices.

### A. *Adaptive Transformers architectures: TS-Forecasting*

Although the Transformer models work well in many applications such as machine translation, text generation, computer vision, there are still some limitations when applying the Transformer-based models to long time-series forecasting problem. Some of the limitations include quadratic time/memory complexity with the original self-attention scheme and error accumulation caused by the autoregressive decoder design (Zeng et al. 2022).

**AutoFormer: Time Series Decomposition**
Building on and extending the Transformer model, Autoformer introduces a novel decomposition architecture with an Auto-Correlation mechanism that replaces traditional series pre-processing. By integrating decomposition as a core part of the model, Autoformer demonstrates state-of-the-art results in long-term forecasting, showing notable improvements across domains such as energy, traffic, and weather (Wu et al. 2022).

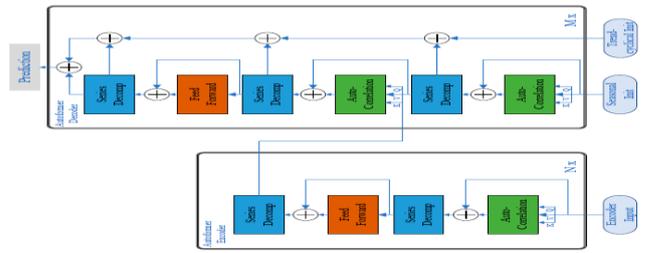

*Figure 1. Autoformer algorithm model architecture (Wu et al., 2022).*

The encoder removes the long-term and cyclical components through a series of decomposition blocks (indicated by blue blocks), enabling it to focus on modelling the seasonal patterns. The decoder accumulates the trend information extracted from hidden variables. The encoder-decoder auto-correlation mechanism (represented by the central green block in the decoder) utilizes past seasonal signals from the encoder to enhance forecasting accuracy (Wu et al. 2022).
In time series modelling, usually the raw data goes through some transformations such as normalization where data has zero mean and unit variance. The Autoformer uses a simple seasonal-trend decomposition architecture with an auto-correlation mechanism working as attention module (Wen et al. 2023). Each block in the Autoformer architecture applies seasonal-trend decomposition to separate the trend, seasonality, and the noise of the raw data to improve the predictability of the data for forecasting tasks.
The encoder section in figure 1, focuses on the seasonal part of the model. The past seasonal information appears in the output part of the encoder. This information is then used as the cross information to help the decoder refine prediction results. The decoder part consists of two sections – trend cyclical component and an Auto-Correlation mechanism for seasonal components (Figure 1). There are two Auto-Correlation layers inside the decoder which utilizes past seasonal information and helps generate revised predictions. The model also extracts trend information from the decoder's hidden layers, allowing Autoformer to gradually improve trend forecasts and filter out noise when identifying periodic patterns. The Autoformer architecture applies a Fast Fourier Transform to compute the autocorrelation.

**Informer: ProbSparse Self-Attention**
Zhou et al. (2023) designed and implemented the Informer model—a refined Transformer architecture—by introducing a probabilistic sparse multi-head self-attention mechanism (ProbSparse). This modification to the original self-attention framework enables the model to selectively focus on the most informative query-key pairs, significantly reducing the time and memory complexity to O (L log$^{-1}$ L), where *L* denotes sequence length. To further improve computational efficiency, Informer applies a self-attention distilling mechanism, which compresses the input sequence across encoder layers, allowing the model to retain essential information while discarding redundancies. Additionally, Informer utilizes a generative-style decoder, which departs from traditional autoregressive models by predicting the entire output sequence in a single forward pass, thereby improving inference speed and minimizing error accumulation.

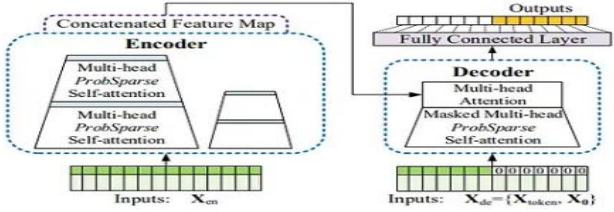

*Figure 2. Informer algorithm model architecture (Zhou et al. 2023)*

These innovations collectively enable Informer to efficiently handle long sequence forecasting tasks across various domains. However, despite these advancements, Zhu et al. (2023) point out that there remains potential for further optimization. While ProbSparse self-attention significantly enhances efficiency, there is still room to reduce computational cost and memory usage. Moreover, challenges with modeling forward and backward temporal dependencies may lead to inconsistencies between predicted and actual values, affecting forecasting accuracy. Finally, the decoder architecture remains relatively simple and closely aligned with that of the original Transformer, suggesting that further architectural enhancements could improve the model's expressiveness and performance.

### B. Adaptive Linear architectures: TS-Forecasting

Despite the recent surge in research on Transformer-based models for long term time-series forecasting, Zeng et al. (2022) critically examine the effectiveness of this research direction. They argue that while Transformers are widely regarded as effective for capturing semantic correlations in long sequences, their performance in time series forecasting is less convincing. Specifically, their study shows that most Transformer-based models fail to extract meaningful temporal dependencies from long input sequences. Contrary to expectations, forecasting errors do not decrease with longer look-back windows and, in some cases, even increase.

Zeng et al. (2022) hypothesize that long-term forecasting is only feasible for time series with clearly defined trends or periodic patterns. Since simple linear models can already capture such characteristics, they introduce LTSF-Linear, a set of embarrassingly simple models that use a one-layer linear regression to directly predict future values from historical observations.

Through extensive experiments across nine widely-used benchmark datasets—including traffic, energy, economics, weather, and disease prediction—they find that LTSF-Linear outperforms complex Transformer-based models in all cases, often by a substantial margin (20%–50%). These findings call into question the current direction of research that prioritizes increasingly complex Transformer architectures for time series forecasting.

**LTSF-Linear Model Definition:**
The most basic form of time series forecasting is based on simple Linear Network which is essentially a regression mode as presented by Zeng et al. (2022), Su (2022), Rizvi et al. (2025), Toner and Darlow (2024). The core approach of Long Time Series Forecasting – Linear involves predicting future time series by directly applying a weighted sum to historical data, as shown in figure 3 (Alharti and Mahmood 2024). Zeng et al. (2022) used a direct multistep model via temporal linear layer, named LSTF-Linear (also known as NLinear), as a baseline for comparison.

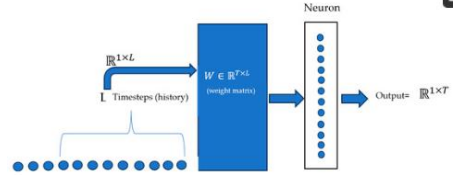

*Figure 3. Predicting T future time-steps based on L time-steps*

The basic formulation of LSTF-Linear directly regresses historical time series for future predictions via a weighted sum operation. The output of LSTF-Linear is described as $\hat{X} = WX_i$ where $W \in R^{T \times L}$ is a temporal linear layer and $X_i$ is the input to the *i-th* variable (Alharti and Mahmood 2024). This simplistic architecture can be visualized in figure 4.

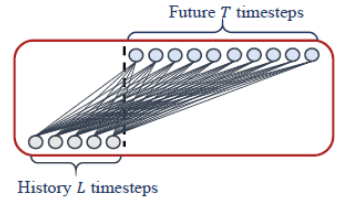

*Figure 4. LSTF-Linear architecture (Zeng et al. 2022).*

The LSTF-Linear (NLinear) shares weights across different variates and does not model any spatial correlations. Due to its inability to model non-linear relationships, this approach tends to underperform in forecasting scenarios involving complex data dynamics. Beside this simplistic baseline Linear architecture, other variations such as RLinear and DLinear were introduced by Zeng et al. (2022).

### RLinear Predictor Model (Reversible normalization-based Linear model- RevIN)

Non stationarity in time series data is a common problem that is present in time series models. This means that statistical properties of time series data such as mean, and variance change over time—this is known as distribution shift problem. However, when a normalization such as removing non-stationary information from the data is applied, it can lead to poor forecasting accuracy. Therefore, if the removed non-stationary information via the normalization is returned back to the model, the model will not have to rebuild the original distribution by itself while keeping the advantage of normalizing the input (Kim et al. 2022). The authors proposed a new normalization-and-denormalization method (RevIN). This method first normalizes the input sequence and then de-normalize the model output sequence to solve time-series forecasting problems against distribution shift. This simply means that the RevIN approaches the distribution discrepancy problems in time-series forecasting by removing non-stationary information in the input layer and then restoring it in the output layer. Li et al. (2023) showed that a simple linear model using RevIN is able to outperform most deep models on standard dataset.

The following is the mathematical formulation of the RevIN normalization-de-normalization method: Given input vector $\vec{x}$ and target output vector $\vec{y}$, the Reversible Instance Normalization (RevIN) for each of the data instance involves a two-step normalization process. First, $\vec{x}$ is normalized by its mean $\mu(\vec{x})$ and its standard deviation $\sigma(\vec{x})$. Next, affine transformation is applied using parameters $\alpha$ and $\beta$. The transformed data $x''$ is then passed through the forecasting model function. This process is then reversed to retrieve the prediction in the original scale (Darlow et al., 2024). Following the derivation by Darlow et al. (2024), This is expressed as:

$$\vec{x'} = \frac{\vec{x} - \mu(\vec{x})}{\sigma(\vec{x}) + \varepsilon}$$

$$\vec{x''} = \frac{\vec{x'} - \beta}{\alpha}$$

$$\hat{y} = f(\vec{x''})$$

$$\widehat{y'} = \alpha \hat{y} + \beta$$

$$\hat{y}_{out} = (\sigma(\vec{x}) + \varepsilon) + \mu(\vec{x})$$

The mean and standard deviation are computed for every instance.

$$\mu(\vec{x}) = \frac{1}{T}\sum_{j=1}^{T} x_{kj}^i \text{ and } \sigma(\vec{x}) = \frac{1}{T}\sum_{j=1}^{T}\left(x_{kj}^i - \mu(\vec{x})\right)^2$$

The alpha and beta values are learnable affine parameter vectors.

This normalization, de-normalization layer proves to be a key component in the revised GLinear model, serving as essential feature in the architecture of our newly proposed RS-GLinear model.

*DLinear Predictor Model (Decomposition-based Linear model)*
DLinear model incorporates the decomposition technique introduced in the Autoformer architecture, applying a moving average kernel on the input series to isolate the trend-cyclical component of the time-series. The seasonal component is the difference after removing the original times series data from its trend. The DLinear model decomposes a time-series data into a trend component $X_t$ and seasonal component $X_s = X - X_t$. Two linear layers are then applied to these two series (Chaoqun Su. 2022).

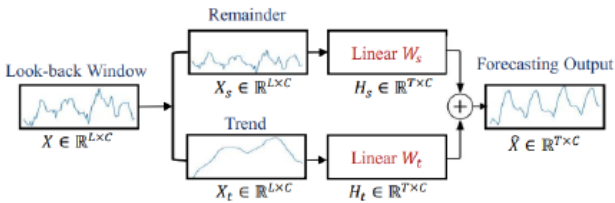

Figure 5. DLinear architecture: (Chaoqun Su., 2022).

We formulate the DLinear process as: $\hat{X} = H_s + H_t$ where $H_s = W_s X_s$ and $H_t = W_t X_t$ are the decomposed trend and remainder features. $W_s$ and $W_t$ are two linear layers applied to each component, and the two features are summed up to get the final prediction (Chaoqun Su. 2022). This feature improves the performance of vanilla linear when there is a clear trend in the data. However, DLinear can be computationally demanding and may often require substantial amount of training data, which may limit its suitability for real-time application and hinder scalability (Rizvi et al. 2025).

*GLinear Predictor Model (Decomposition-based Linear model)*
Due to the complexity and overfitting problems in Transformer-based architecture for time-series modelling, Rizvi et al. (2025) investigated how to effectively integrate the simplicity of linear models with sophisticated techniques for capturing complex underlying patterns to further enhance medium and long-term time-series forecasting. While retaining the simplicity of the above Linear Models, they proposed a novel and data efficient architecture named GLinear (Gaussian-based linear predictor) that does not include self-attention blocks from the Transformer models architecture. However, the model introduces a new modification to the linear architecture by integrating two additional layers. It adds a non-linear Gaussian Error Linear Unit (GeLU) under its transformation layer to capture intricate patterns. The second modification in its architecture (figure 6-a) involves a new Reversible Instance Normalization layer (RevIN) that allows standardization of the data distribution across instances, ensuring consistency in performance and adaptability across diverse datasets (Rizvi et al. 2025).

III. PRELIMINARIES: TSF PROBLEM FORMULATION

Consider a time series data consisting of N variates with historical data $\{\chi = X_1^t, ... X_N^t\}_{t=1}^{L}$, where $L$ is the look-back window size and $X_1^t$ is the *i-th* variates at the *t-th* time-step. The goal is to predict the values $\{\chi = X_1^t, ... X_N^t\}_{t=L+1}^{L+T}$ at the $T$ future time-steps (Zeng et al. 2022). Periods when $T > 1$, there are either a single-step or multiple-step predictions that can be applied (Iterated Multi-Step forecasting and Direct Multi-step forecasting) (Taieb et al. 2012, Chevillon. 2007)

IV. METHODOLOGY

RESIDUAL-STACKED GLINEAR MODEL ARCHITECTURE

The baseline GLinear model architecture is composed of two fully connected layers with the same input dimension. A non-linear Gaussian Error Linear Unit (GeLU) [2] is added in between the two fully connected layers. The non-linear activation function GeLU is defined as:

$$GeLU(x) = x \cdot \Phi(x)$$

Where $\Phi(x)$ is the cumulative distribution function of the standard normal distribution with an error function (*erf)*, and $x$ is the input data in the neural network.

---

[2] *Mathematical details, see research paper by Lee, M. (2023), Hendrycks et al. (2023)*

$$\Phi(x) = \frac{1}{2}\left(1 + erf\frac{x}{\sqrt{2}}\right)$$

For practical implementation purposes in the neural network, the following approximation is usually used:

$$0.5x\left(1 + \tanh\left[\sqrt{\frac{2}{\pi}}(x + 0.04471x^3)\right]\right)$$

To reduce distribution related discrepancies between the input and the output data, Reversible Instance (RevIN) Normalization and Denormalization are also applied. These normalization layers are added at the start and the end of the GLinear architecture in order to allow the data to be centered around its mean. Therefore, this normalization allows the GLinear model to handle shifts in the temporal distribution over time. Building upon this framework, we adopt the original GLinear architecture and introduce—**Residual-Stacked GLinear** Model (Depth =4), which is an enhanced version of the original GLinear architecture that integrates more depth in the neural network via stacked linear blocks (figure 6-b). Our proposed enhanced GLinear architecture retains the original Reversible Instance activation function. As we increase the network linear layers or build deeper neural networks, there is a risk that our network may also suffer from the so-called "degradation problem" as described by He et al. (2015), where deeper layers may lead to higher training error due to vanishing or exploding gradients. To resolve the "degradation problem", we introduce residual block and skip connection. The skip connection in a residual block allows the input to bypass intermediate layers and directly connect a shallow layer to a deeper one. He et al. (2015) provides the following mathematical representation for the skip mechanism:

$$h(x) = \mathcal{F}(x) + x$$

Where $\mathcal{F}(x)$ represents the function learned by the layers or linear transformations from relevant layers (Residuals). While the positive $x$ allows the original input to pass through. Therefore, $h(x)$ becomes identity function if the residue $\mathcal{F}(x)$ becomes zero as a result of zero gradient such that.

$$h(x) = 0 + x$$

which is the necessary identity function. This helps address the degradation problem

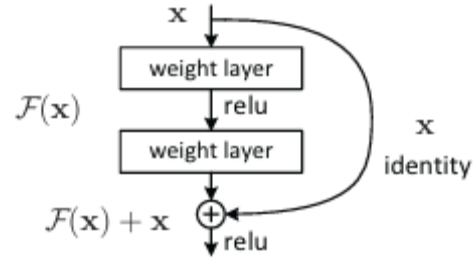

*Figure 6-c. Residual learning a building block (He et al. 2015).*

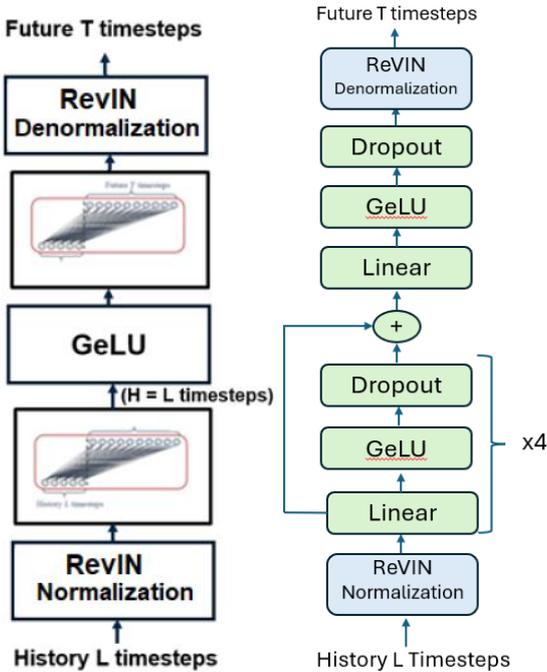

*Figure 6-a. original GLinear architecture (left) (Rizvi et al 2025). Figure 6-b. Architecture of the proposed RS-GLinear model with stacked linear blocks and residual connections (right)*

Normalization (RevIN) components—used before input and output layers—but replaces the shallow linear layer with sequence of stacked linear transformations, new residual skip connections and dropout layers. The stacked blocks consist of fully connected Linear layers each followed by an activation function (GeLU) and Dropout layers that are repeated (x4) with residual skip connection feeding back into the next block (figure 6-b). These stacked blocks are essentially residual blocks, a common architectural element in neural networks. A residual block typically consists of convolution layer, followed by batch normalization, and an

## V. Experiments

In this section, we present an empirical study to evaluate and analyze our proposed Residual-Stacked GLinear (RS-GLinear) model. First, we provide a summary of the datasets and experimental setting. We will then compare our proposed framework with existing multivariate time-series forecasting methods on the benchmark datasets.

*A. Multivariate Time-Series Datasets.*

Our experiments are conducted using six multivariate time-series datasets, each with distinct characteristics and widely adopted in time-series forecasting research. These datasets have been utilized in several influential studies, including those by Zeng et al. (2022), Zhu et al. (2023), Zhou et al. (2021), Wu et al. (2022), Rizvi et al. (2025), and Tan et al. (2024). Owing to their standard nature, these datasets are often accompanied by well-established descriptive approaches across the literature. In alignment with Zeng et al. (2022), we adopt a similar methodology to describe and evaluate these datasets in the context of our work.

**(1) ETTh1** (Electricity Transformer Temperature) contains data collected from electricity transformers, including load oil temperature that are recorded every hour between July 2016 to July 2018 from two different regions in China. The dataset consists of seven features, including power load features and oil temperature- crucial indicator data in the electric power long-term deployment. The ETTh1 is a multivariate time-series dataset, and the sample size used for this study consists of 17420 data points, 7 features and hourly sampled dataset. **(2) Traffic** is a collection of hourly data from California Department of Transportation, which describes the road occupancy rates measured by different sensors on San Francisco Bay area freeways. The data consists of 17544 data points, 862 features with 1-hour granularity. **(3) Weather** dataset contains more frequent granularity, with measurements recorded every 10 minutes in Germany during the year 2020. This dataset consists of 52696 data points and 21 features (weather indicators). **(4) Electricity** dataset is based on the hourly electricity consumption of 321 clients from 1990 to 2016. The data consists of 26304 data points, 321 features with 1-hour granularity. **(5) Exchange Rate** consists of collection of daily exchange rates of 8 countries from 1990 to 2016. The data consists of 7588 data points, 8 currency pairs (each country) against the USD and end-of-day observations (granularity). **(6) National Illness** describes the ratio of patients seen with influenza-like symptoms and the number of patients. It includes weekly data from Centers for Disease Control (CDC) and Prevention in USA from 2002 to 2021.

We follow standard protocol and split all datasets into training, validation and test set in chronological order by the ratio of 6:2:2. Table 1 shows summary of the datasets. Figure 7 shows the size of the time-series dataset expressed in log-scale.

*Table 1: Statistics of popular datasets for benchmark*

| Data Type | TS-Length | Nr of Variables | Size - data points | Scaled (Mln) | Sample Rate |
|---|---|---|---|---|---|
| ETTh1 | 17420 | 7 | 121,940.00 | 0.12 | 1 hour |
| Electricity | 26304 | 321 | 8,443,584.00 | 8.44 | 1 hour |
| Traffic | 17544 | 862 | 15,122,928.00 | 15.12 | 1 hour |
| Weather | 52696 | 21 | 1,106,616.00 | 1.11 | 10 minutes |
| Ex-Rate | 7588 | 8 | 60,704.00 | 0.06 | 1 day |
| National Illness | 966 | 7 | 6762 | 0.006762 | weekly |

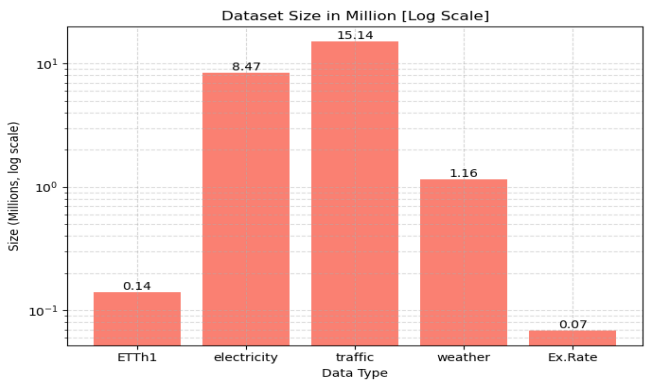

*Figure 7. Time-series data size by type in Million.*

### B. Implementation details

Our implementation is built on the PyTorch framework and adapts to the baseline GLinear code introduced by Rizvi et al. (2025). Zeng et al. (2022) provide the baseline codes for key Transformer inspired implementations (Informer, Autoformer) as well as LSTF-Linear models (DLinear, NLinear). We conducted our experiment on a single A100 GPU. We used a batch size of 32 and trained each model for up to 10 epochs using Adam optimizer, which incorporated early stopping to prevent overfitting. We used a learning rate of 0.001 (default setting for the GLinear model across all datasets) for most of the datasets except for ETTh1 dataset. In order to achieve a lower forecasting error rate (MSE and MAE), we fine-tuned the learning to 0.01 for ETTh1 dataset.

### C. Evaluation Details

To evaluate the performance of our implementations, we adopted two widely used loss functions in time-series forecasting literature: Mean Squared Error (MSE) and Mean Absolute Error (MAE). MSE measures the average squared error difference between the predicted values and actual observations. The MSE ensures that the larger errors are magnified by the squared term and therefore the minimization process is accelerated during training. The MAE calculates the average of the absolute differences which makes it inherently robust to outliers. These metrics are formally defined as:

$$MSE = \frac{1}{m}\sum_{i=1}^{m}(y_i - \hat{y}_i)^2$$

$$MAE = \frac{1}{m}\sum_{i=1}^{m}|y_i - \hat{y}_i|$$

Where $y_i$ denotes the actual observations, $\hat{y}_i$ the predicted values, and $m$ the sample size. Our goal is to minimize both MSE and MAE across different models and datasets. A lower value in MSE or MAE metrics indicates better predictive accuracy. By comparing these results, we aim to demonstrate that the proposed RS-GLinear model achieves an improved MSE, MAE accuracy measure, especially in cases where the selected baseline model (GLinear) showed poor performances against both linear models and Transformer based models (see table 2-c).

## VI. EXPERIMENTS AND RESULTS

We divide our experiment results into two parts. The first set of results focuses on the performance of our proposed RS-GLinear model relative to the baseline (benchmark) GLinear model using ETTh1, Electricity, Weather and Traffic datasets described in the study by Rizvi et al. (2025). We fixed the input sequence of our experiment to 336 time-steps and evaluated model performance across multiple forecast horizons: {12, 24, 48, 96, 192, 336, 720}[3]. As shown in Tables 2-a and 2-c, the RS-GLinear model consistently outperforms (showing lower MSE, MAE errors) not only the benchmark GLinear model but also Transformer based time-series forecasting (Autoformer)[4] and Linear models in most cases, achieving lower MSE and MAE values. Specially, for the electricity dataset, RS-GLinear achieves additional performance gains over GLinear, reducing error ranging approximately from 0.6% to 5.5% in MSE and 0.1% to 3% in MAE respectively across all forecast horizons. The only exception occurs when the prediction length equals the fixed input length of 336 time-steps, where no significant gain is observed. The underperformance of the RS-GLinear model at prediction length of 336—equal to the fixed input sequence length—is consistently observed across all four primary datasets presented in table 2-a. This suggests potential limitation in the model's capacity to perform well when the input window and the forecasting horizon match.

---

[3] *Given the granularity of the data in table 1, the input sequence and the set of time frames can be translated into days.*

[4] *Autoformer as Transformer-based benchmark model using datasets: Electricity, ETTh1, Traffic and Weather datasets.*

Table 2-a: RS-GLinear vs GLinear   Table 2-b: Performance in %

| Dataset | Pred.Length | GLinear MSE | GLinear MAE | RS-GLinear MSE | RS-GLinear MAE |
|---|---|---|---|---|---|
| Electricity | 12 | 0.0883 | 0.1860 | 0.0836 | 0.1802 |
| | 24 | 0.0988 | 0.1952 | 0.0949 | 0.1910 |
| | 48 | 0.1144 | 0.2101 | 0.1116 | 0.2069 |
| | 96 | 0.1313 | 0.2258 | 0.1299 | 0.2241 |
| | 192 | 0.1494 | 0.2423 | 0.1495 | 0.2414 |
| | 336 | 0.1651 | 0.2582 | 0.1788 | 0.2699 |
| | 720 | 0.2027 | 0.2906 | 0.2013 | 0.2903 |
| ETTh1 | 12 | 0.2848 | 0.3448 | 0.2973 | 0.3489 |
| | 24 | 0.3142 | 0.3654 | 0.3206 | 0.3644 |
| | 48 | 0.3537 | 0.3869 | 0.3420 | 0.3761 |
| | 96 | 0.3820 | 0.4025 | 0.3701 | 0.3914 |
| | 192 | 0.4202 | 0.4269 | 0.4059 | 0.4137 |
| | 336 | 0.4915 | 0.4715 | 0.5122 | 0.4989 |
| | 720 | 0.5923 | 0.5372 | 0.4543 | 0.4632 |
| Traffic | 12 | 0.3222 | 0.2385 | 0.3070 | 0.2269 |
| | 24 | 0.3369 | 0.2471 | 0.3229 | 0.2357 |
| | 48 | 0.3630 | 0.2607 | 0.3477 | 0.2479 |
| | 96 | 0.3875 | 0.2718 | 0.3760 | 0.2619 |
| | 192 | 0.4056 | 0.2802 | 0.3960 | 0.2701 |
| | 336 | 0.4200 | 0.2871 | 0.4523 | 0.2922 |
| | 720 | 0.4488 | 0.3038 | 0.4393 | 0.2949 |
| Weather | 12 | 0.0716 | 0.0940 | 0.0737 | 0.1045 |
| | 24 | 0.0909 | 0.1247 | 0.0940 | 0.1206 |
| | 48 | 0.1163 | 0.1602 | 0.1193 | 0.1650 |
| | 96 | 0.1457 | 0.1966 | 0.1514 | 0.2010 |
| | 192 | 0.1883 | 0.2385 | 0.1953 | 0.2451 |
| | 336 | 0.2407 | 0.2764 | 0.2961 | 0.3118 |
| | 720 | 0.3200 | 0.3334 | 0.3241 | 0.3358 |

| Dataset | Pred.Length | %diff MSE | %diff MAE |
|---|---|---|---|
| Electricity | 12 | -5.53% | -3.17% |
| | 24 | -4.03% | -2.18% |
| | 48 | -2.48% | -1.53% |
| | 96 | -1.07% | -0.76% |
| | 192 | 0.07% | -0.36% |
| | 336 | 7.97% | 4.43% |
| | 720 | -0.69% | -0.10% |
| ETTh1 | 12 | 4.30% | 1.18% |
| | 24 | 2.00% | -0.27% |
| | 48 | -3.36% | -2.83% |
| | 96 | -3.16% | -2.80% |
| | 192 | -3.46% | -3.14% |
| | 336 | 4.13% | 5.65% |
| | 720 | -26.53% | -14.82% |
| Traffic | 12 | -0.0483 | -0.0499 |
| | 24 | -0.0424 | -0.0472 |
| | 48 | -0.0431 | -0.0503 |
| | 96 | -0.0301 | -0.0371 |
| | 192 | -0.0240 | -0.0367 |
| | 336 | 0.0741 | 0.0176 |
| | 720 | -0.0214 | -0.0297 |
| Weather | 12 | 0.0284 | 0.1059 |
| | 24 | 0.0335 | -0.0334 |
| | 48 | 0.0255 | 0.0295 |
| | 96 | 0.0384 | 0.0221 |
| | 192 | 0.0365 | 0.0273 |
| | 336 | 0.2071 | 0.1205 |
| | 720 | 0.0127 | 0.0072 |

*Table 2-a: RS-GLinear vs GLinear results with prediction length {12, 24, 48, 96, 192,....720}. Input length I as 96. Lower MAE or MSE indicate a better prediction. Table 2-b: GLinear vs RS-GLinear performance comparison. Negative value indicates RS-GLinear provides further error reduction.*

Furthermore, on the weather dataset, the RS-GLinear model does not demonstrate any significant improvement over the baseline GLinear model, with performance remaining largely comparable (see figure 8). However, RS-GLinear still achieves lower prediction errors compared to the Transformer-based model on the weather dataset.

On the ETTh1 dataset, the RS-GLinear model achieves similar improvements over the baseline GLinear model, with average error reduction of approximately 3% in both MSE and MAE across most forecasting horizons. At the longest forecasting horizon (720 time-steps), the RS-GLinear delivers substantial reduction in error, lowering MSE by 26.5% and MAE by 14.7%. This suggests that the enhanced linear model has great capacity to handle long-range temporal dependencies. Improvements over the GLinear model on the Traffic dataset are consistent across all forecasting horizons; however, the error reduction is very marginal, and at best, the RS-GLinear performs on par with the GLinear model.

*Table 2-c: Mult. time-series errors across models in terms of MSE and MAE*

| Methods | | Lookup Window (Input Sequence Length) = 336 Learning Rate 0.001 | | | | | | | | |
|---|---|---|---|---|---|---|---|---|---|---|
| | | Autoformer | | NLinear | | DLinear | | RLinear | | GLinear | |
| Dataset / Output Horizon (Prediction Length) | | MSE | MAE | MSE | MAE | MSE | MAE | MSE | MAE | MSE | MAE |
| Electricity | 12 | 0.1638 | 0.2872 | 0.1000 | 0.2006 | 0.0997 | 0.2009 | 0.0967 | 0.1973 | 0.0883 | 0.1860 |
| | 24 | 0.1711 | 0.2917 | 0.1103 | 0.2092 | 0.1099 | 0.2089 | 0.1072 | 0.2049 | 0.0988 | 0.1952 |
| | 48 | 0.1827 | 0.2990 | 0.1255 | 0.2232 | 0.1249 | 0.2231 | 0.1201 | 0.2180 | 0.1144 | 0.2101 |
| | 96 | 0.1960 | 0.3106 | 0.1409 | 0.2366 | 0.1401 | 0.2374 | 0.1358 | 0.2317 | 0.1313 | 0.2258 |
| | 192 | 0.2064 | 0.3182 | 0.1551 | 0.2488 | 0.1538 | 0.2505 | 0.1518 | 0.2455 | 0.1494 | 0.2423 |
| | 336 | 0.2177 | 0.3290 | 0.1717 | 0.2654 | 0.1693 | 0.2678 | 0.1688 | 0.2621 | 0.1651 | 0.2582 |
| | 720 | 0.2477 | 0.3528 | 0.2104 | 0.2977 | 0.2042 | 0.3005 | 0.2071 | 0.2940 | 0.2027 | 0.2906 |
| ETTh1 | 12 | 0.3991 | 0.4422 | 0.3069 | 0.3564 | 0.2976 | 0.3494 | 0.2862 | 0.3412 | 0.2848 | 0.3448 |
| | 24 | 0.4759 | 0.4733 | 0.3474 | 0.3842 | 0.3194 | 0.3627 | 0.3090 | 0.3559 | 0.3142 | 0.3654 |
| | 48 | 0.5046 | 0.4831 | 0.3553 | 0.3845 | 0.3477 | 0.3803 | 0.3454 | 0.3766 | 0.3537 | 0.3869 |
| | 96 | 0.5392 | 0.4979 | 0.3731 | 0.3941 | 0.3705 | 0.3919 | 0.3901 | 0.4054 | 0.3820 | 0.4025 |
| | 192 | 0.4907 | 0.4906 | 0.4089 | 0.4157 | 0.4044 | 0.4128 | 0.4223 | 0.4279 | 0.4202 | 0.4269 |
| | 336 | 0.4805 | 0.4886 | 0.4324 | 0.4307 | 0.4553 | 0.4582 | 0.4417 | 0.4383 | 0.4915 | 0.4715 |
| | 720 | 0.6303 | 0.5930 | 0.4369 | 0.4527 | 0.4975 | 0.5087 | 0.4634 | 0.4686 | 0.5923 | 0.5372 |
| Traffic | 12 | 0.5624 | 0.3830 | 0.3623 | 0.2662 | 0.3610 | 0.2644 | 0.3762 | 0.2744 | 0.3222 | 0.2385 |
| | 24 | 0.5801 | 0.3786 | 0.3719 | 0.2682 | 0.3709 | 0.2672 | 0.3834 | 0.2775 | 0.3369 | 0.2471 |
| | 48 | 0.6060 | 0.3796 | 0.3945 | 0.2769 | 0.3932 | 0.2760 | 0.4041 | 0.2864 | 0.3630 | 0.2607 |
| | 96 | 0.6426 | 0.3998 | 0.4113 | 0.2820 | 0.4104 | 0.2829 | 0.4194 | 0.2921 | 0.3875 | 0.2718 |
| | 192 | 0.6425 | 0.3967 | 0.4245 | 0.2872 | 0.4229 | 0.2881 | 0.4323 | 0.2965 | 0.4056 | 0.2802 |
| | 336 | 0.6675 | 0.4088 | 0.4375 | 0.2943 | 0.4362 | 0.2961 | 0.4451 | 0.3027 | 0.4200 | 0.2871 |
| | 720 | 0.6570 | 0.4030 | 0.4657 | 0.3109 | 0.4660 | 0.3152 | 0.4733 | 0.3191 | 0.4488 | 0.3038 |
| Weather | 12 | 0.2010 | 0.2933 | 0.0784 | 0.1127 | 0.0783 | 0.1158 | 0.0706 | 0.0974 | 0.0716 | 0.0940 |
| | 24 | 0.2095 | 0.3033 | 0.1056 | 0.1453 | 0.1040 | 0.1519 | 0.0905 | 0.1247 | 0.0909 | 0.1247 |
| | 48 | 0.2397 | 0.3202 | 0.1357 | 0.1824 | 0.1367 | 0.1937 | 0.1138 | 0.1566 | 0.1163 | 0.1602 |
| | 96 | 0.3004 | 0.3776 | 0.1761 | 0.2264 | 0.1756 | 0.2386 | 0.1450 | 0.1936 | 0.1457 | 0.1966 |
| | 192 | 0.3916 | 0.4382 | 0.2164 | 0.2595 | 0.2160 | 0.2739 | 0.1878 | 0.2339 | 0.1883 | 0.2385 |
| | 336 | 0.3830 | 0.4171 | 0.2664 | 0.2966 | 0.2652 | 0.3192 | 0.2404 | 0.2743 | 0.2407 | 0.2764 |
| | 720 | 0.5420 | 0.5032 | 0.3339 | 0.3437 | 0.3275 | 0.3667 | 0.3159 | 0.3271 | 0.3200 | 0.3334 |

*Source: Rizvi et al. (2025). *Multivariate*

The second set of results evaluates the performance of our proposed RS-GLinear model relative to several Transformer-based models, such as **Fedformer**, **Autoformer**, **Informer**, **Pyraformer** as well as best performing **NLinear** model, on the National Illness (ILI) dataset – a dataset that were not included in the original benchmark study by Rizvi et al. (2025). The performance metrics of these Transformer-based models on ILI data across multiple forecast horizons were previously reported by Zeng et al. (2022). We trained the RS-GLinear model on the National Illness (ILI) dataset using an input window of 96 against and forecast horizons of: {24, 36, 48, 60}. The result demonstrates that the RS-GLinear model still outperforms the baseline Transformer-based models while NLinear still remains the best-performing model. However, although the simpler linear model (NLinear) remains the overall best-performing model, RS-GLinear achieves the lowest error (both MSE and MAE) at the longest forecast horizon of 60. Table 3-a highlights the best result in bold along with their corresponding best model. Figure 8 shows the RS-GLinear performances vs baseline model (GLinear).

*Table 3-a: Forecasting errors across models based on MSE and MAE*

| | | NLinear | | FEDformer | | AutoFormer | | Informer | | Pyraformer | | RS-GLinear | |
|---|---|---|---|---|---|---|---|---|---|---|---|---|---|
| Dataset | Pred.Length | MSE | MAE | MSE | MAE | MSE | MAE | MSE | MAE | MSE | MAE | MSE | MAE |
| National Illness | 24 | 1.6830 | 0.8580 | 3.2280 | 1.2600 | 3.4830 | 1.2870 | 5.7640 | 1.6770 | 1.4200 | 2.0120 | 1.7960 | 0.8484 |
| | 36 | 1.7030 | 0.8590 | 2.6790 | 1.0800 | 3.1030 | 1.1480 | 4.7550 | 1.4670 | 7.3940 | 2.0310 | 1.8496 | 0.8871 |
| | 48 | 1.7190 | 0.8840 | 2.6220 | 1.0780 | 2.6690 | 1.0850 | 4.7630 | 1.4690 | 7.5510 | 2.0570 | 1.7942 | 0.9001 |
| | 60 | 1.8190 | 0.9170 | 2.8570 | 1.1570 | 2.7700 | 1.1250 | 5.2640 | 1.5640 | 7.6620 | 2.1000 | 1.7728 | 0.9040 |

As part of our second and final set of results, we evaluate the performance of our proposed RS-GLinear model relative to several Transformer-based models, including **Autoformer**, **Informer**, **Logtransform**, **Reformer** on exchange rate dataset. This dataset was also not included in the original benchmark study by Rizvi et al. (2025). The performance metrics of these Transformer-based models on exchange rate data across multiple forecast horizons were previously reported by Wu et al. (2022). We trained the RS-GLinear model on the foreign exchange rate dataset using an input window of 96 time-steps and forecast horizons of {96, 192, 336, 720}. The RS-GLinear model achieved a substantial reduction in prediction errors compared to all the Transformer-based models on the foreign exchange dataset. Specially, RS-GLinear model outperformed competing models with error reductions ranging from -238% to -40% in terms of MSE, and -132% to -21% in terms of MAE across all forecast horizons. Table 3-b shows the RS-GLinear achievement in lowest MSE and MAE respectively while Table 3-c reports the percentage reduction in errors.

*Table 3-b: Mult. time-series errors across models in terms of MSE and MAE*

| *Input sequence = 96, Learning Rate = 0.01 | | | | | | | | | | |
|---|---|---|---|---|---|---|---|---|---|---|
| Model Type | | AutoFormer | | InFormer | | LogTransForm | | ReFormer | | *RS-GLinear | |
| Dataset | Pred.Length | MSE | MAE | MSE | MAE | MSE | MAE | MSE | MAE | MSE | MAE |
| Exch. Rate | 96 | 0.197 | 0.323 | 0.847 | 0.752 | 0.968 | 0.812 | 1.065 | 0.829 | **0.0985** | **0.2220** |
| | 192 | 0.3 | 0.369 | 1.204 | 0.895 | 1.04 | 0.851 | 1.188 | 0.906 | **0.1782** | **0.2981** |
| | 336 | 0.509 | 0.524 | 1.672 | 1.036 | 1.659 | 1.081 | 1.357 | 0.976 | **0.3420** | **0.4207** |
| | 720 | 1.447 | 0.941 | 2.478 | 1.31 | 1.941 | 1.127 | 1.51 | 1.016 | **0.9445** | **0.7309** |

*Table 3-c: Exchange Rate MSE, MAE reduction in % relative to RS-GLinear*

| *Input sequence = 96, Learning Rate = 0.01 | | | | | | | | | | |
|---|---|---|---|---|---|---|---|---|---|---|
| Model Type | | AutoFormer | | InFormer | | LogTransForm | | ReFormer | | *RS-GLinear | |
| Dataset | Pred.Length | MSE | MAE | MSE | MAE | MSE | MAE | MSE | MAE | MSE | MAE |
| Exch. Rate | 96 | -69% | -37% | -215% | -122% | -229% | -130% | -238% | -132% | **0.0985** | **0.2220** |
| | 192 | -52% | -21% | -191% | -110% | -176% | -105% | -190% | -111% | **0.1782** | **0.2981** |
| | 336 | -40% | -22% | -159% | -90% | -158% | -94% | -138% | -84% | **0.3420** | **0.4207** |
| | 720 | -43% | -25% | -96% | -58% | -72% | -43% | -47% | -33% | **0.9445** | **0.7309** |

The overall performance of both the RS-GLinear and GLinear models on the primary benchmark datasets—Electricity, ETTh1, and Traffic—demonstrates a consistent advantage over more complex Transformer-based architectures that incorporate sophisticated attention mechanisms. This suggests that simpler linear models, when properly optimized, may be more effective for time series data characterized by strong seasonal and periodic patterns. For example, time-series forecasting architectures such as Informer and Autoformer incorporate complex features such as ProbSparse attention (Zhou et al. 2021) and time-series decomposition architecture with auto-correlation mechanism (Wu et al. 2022) to enhance forecasting accuracy in multivariate time-series setting. While these added features in the model architecture may provide improvement in capturing the temporal dependencies, however, their reliance on too many parameters tends to cause overfitting and for this reason, LSTF-Linear models may perform better than Transformer-based models (Zeng et al.2022). According to Alharti. M, Mahmood, A. (2024) on their observations on long-term dependency challenges, they reported that Transformers, despite their theoretical potential, often find it challenging to handle very long sequences typical in time series forecasting, largely due to training complexities and gradient dilution. Zeng et al. (2022) also hypothesized that the size of the look-back window impacts forecasting accuracy, aiming to understand how much of the accuracy can be linked to longer historical input. In their experiments, they tested several look-back window sizes: L ∈ {24, 48, 72, 96, 120, 144, 168, 192, 336, 504, 672, 720}. They found that the performance of Transformer-based models either drops or stays the same as the look-back window increases. In contrast, all LTSF-Linear models showed significant improvement with longer input windows. In our case, we fixed the input window size to 336, which is enough to represent medium to long look-back windows. We did not include experiments on varying look-back sizes due to the additional computational burden required to perform parallel runs of all look-back window time-steps. Our results are consistent with those of Zeng et al. (2022), showing that Transformer-based models perform worse with larger look-back windows, while linear models perform better. Therefore, these solutions tend to overfit temporal noises instead of extracting temporal information if given longer sequence. Furthermore, since the relationship between input sequence length and model performance is often data-dependent (Wu et al. 2022), the fixed input window size used in our experiments may be insufficient for complex datasets such as weather, exchange rate, and ILI. Specially, the absence of seasonality and the noisy characteristics of weather and exchange rate dataset may contribute to the decline in forecasting accuracy.

*Figure 8. RS-GLinear performance across forecasting horizons, datasets against GLinear model*

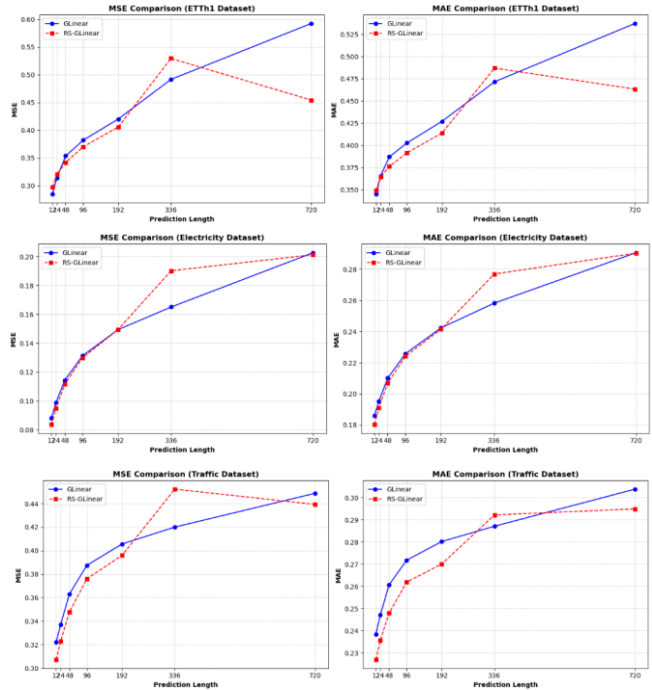

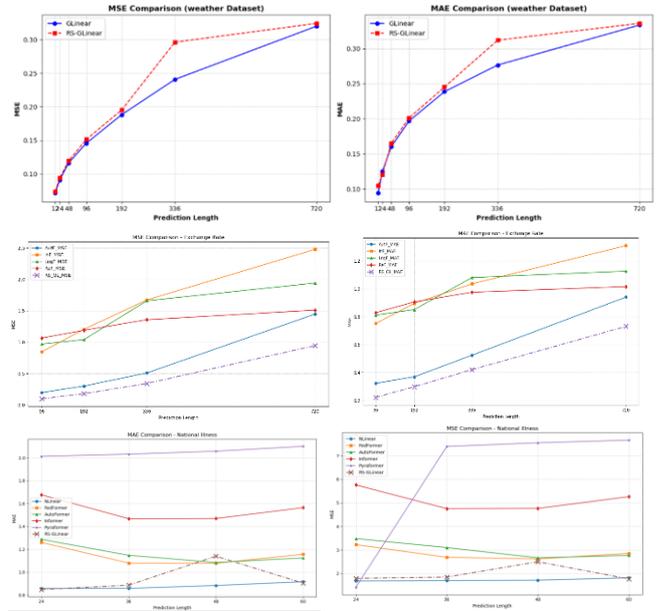

*Figure 9. RS-GLinear performance: Prediction vs Ground-truth*

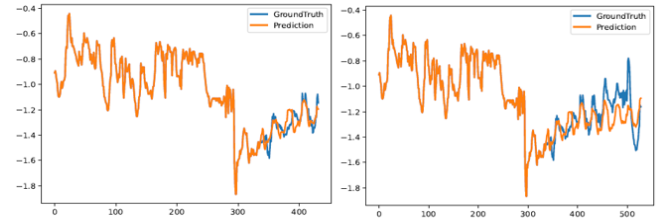

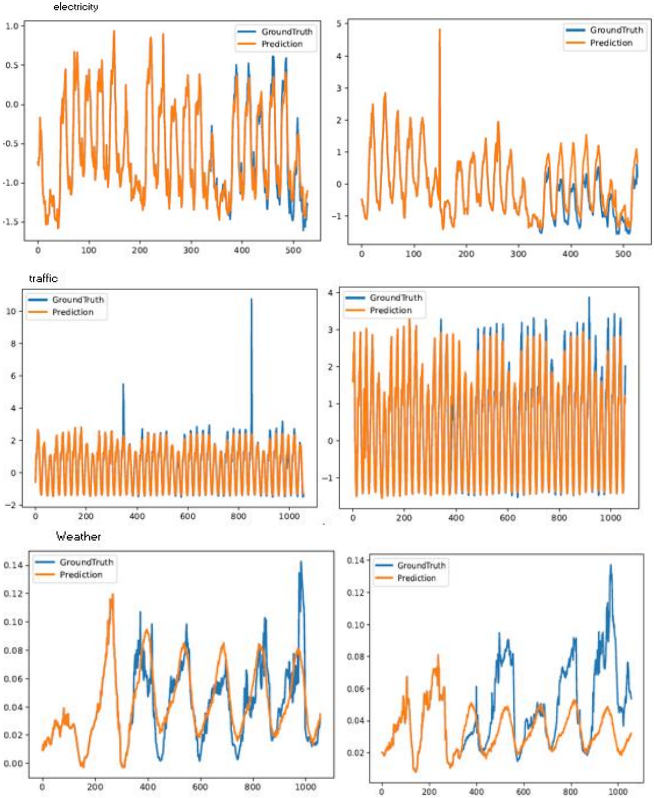

*ETTh1, Electricity and Traffic Datasets, the prediction series seems closer to the ground truth.*

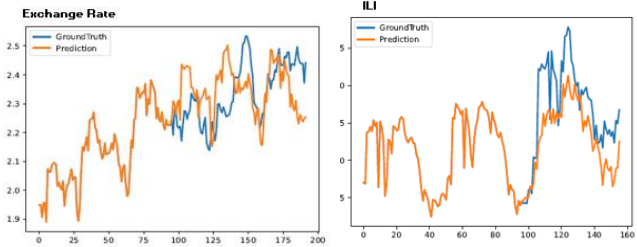

*Exchange and ILI: prediction series deviates from ground truth.*

These results indicate that Transformer-based models and Large Language Models (LLMs) do not necessarily provide superior performance in multivariate time series forecasting tasks solely due to their complex, state-of-the-art architectural designs. These findings are consistent with results reported by other researchers, including Tan et al. (2024) and Zeng et al. (2022).

According to Zeng et al. (2022), the complex positional encoding feature in Transformer architecture is important for text or language modelling but not necessarily for numerical sequences and time-dependent datasets such as time series. Transformer-based models lose essential temporal information when applied to time series data. While these models are effective at capturing semantic correlations, time series data typically lacks such semantics. Instead, it depends on characteristics such as local trends, periodicity, stationarity, stochastic or white noise, and auto-correlations. Although some modifications in Transformer-based architectures attempt to address components like seasonality and trend, they still fail to effectively extract information from noisy time series data that exhibits stochastic behaviour.

From architectural perspective, the findings by Zeng et al. (2022) are particularly compelling. Their study examined whether the self-attention mechanism is effective for long-term time series forecasting tasks—specifically, whether the complex features of Transformer-based models are essential. To validate this, the authors replaced the self-attention layer by a linear layer. Furthermore, they discarded auxiliary design in the Informer network and simplified the model to one Linear layer. Their results were striking, showing that as the Informer model underwent gradual transformation (simplified), its performance improved.

## VII. CONCLUSION AND FUTURE RESEARCH

The purpose of this research was to investigate the validity of simpler, literature-based architectures compared to more complex Transformer-based models for multivariate time-series forecasting tasks. We believe one of the most comprehensive investigations on this topic was conducted by Zeng et al. (2022), who concluded that state-of-the-art Transformer-based models were unable to outperform several linear models. Rizvi et al. (2025) carried out similar investigations and introduced a novel Gaussian-based linear predictor (GLinear) that required less historical data than many leading linear benchmark models.

We adopted the GLinear model and proposed an enhanced architecture—Residual-Stacked GLinear—which increases the depth of the neural network through four stacked linear transformation blocks, each followed by Gaussian Error Linear Unit (GeLU) activations and dropout layers. Additionally, residual skip connections were introduced while retaining the Reversible Instance Normalization (RevIN) layers from the original GLinear architecture.

We tested our model on four multivariate time-series datasets, each with distinct characteristics and widely used in time-series forecasting research: Electricity, Traffic, Weather, and ETTh1. The enhanced RS-GLinear model consistently outperformed the original GLinear in most cases, with the exception of the Weather dataset, where both models performed comparably. Notably, on the ETTh1 dataset, RS-GLinear achieved up to 26.5% and 14.7% improvements in MSE and MAE, respectively, at the longest prediction horizon of 720. These results suggest that the proposed enhancements improve the model's ability to capture long-range temporal dependencies. However, the RS-GLinear model showed consistent underperformance across all primary datasets when the prediction length matched the fixed input length of 336 time-steps. This suggests a potential limitation in the model's capacity to perform well when the input window and forecasting horizon are the same.

In addition to the benchmark datasets, the RS-GLinear model was applied to new domains—financial time series and epidemiological data—to evaluate its generalizability. While the model showed potential, its performance in these domains may be impacted by non-stationary behaviour, the absence of clear seasonality in the exchange rate data, and the presence of noise.

Due to limited computational resources and time constraints, our research restricted the input sequence length to a fixed 336 time-steps while applying short, medium, and long prediction horizons. Further optimization of the model architecture by increasing network depth, additional permutations of features to enhance performance was also beyond the scope of this research due to computational demands. We recommend these areas as potential directions for future work.

Overall, our results align with the findings of Rizvi et al. (2025) and Zeng et al. (2022), which demonstrate that enhanced linear models can produce lower errors than complex Transformer-based models for time series tasks. We conclude that increased complexity in language models does not necessarily offer advantages over simpler neural network models, such as GLinear and RS-GLinear, when these are equipped with architectural modifications that improved learning in multivariate time series forecasting tasks.

VIII. APPENDIX

**Figure A1**. Overview of Transformer-based model architecture as described by Zeng et al. (2022). The pipeline is divided into four main stages: (a) Preprocessing, which includes normalization, seasonal-trend decomposition, and timestamp processing; (b) Embedding, involving global and local timestamp features, positional embeddings, and channel projections; (c) Encoder, which applies a variety of mechanisms such as frequency-enhanced blocks, multi-scale attention, and auto-correlation modules combined with decomposition (e.g., FEDformer, Autoformer, DLinear); and (d) Decoder, which mirrors the encoder in structure and introduces similar decomposition-based modules adapted to specific models. These architectural variations were introduced to address several known limitations of the original Transformer model—specifically, its quadratic time and memory complexity due to the self-attention mechanism, and the risk of error accumulation associated with its autoregressive decoder design (Zeng et al., 2022).

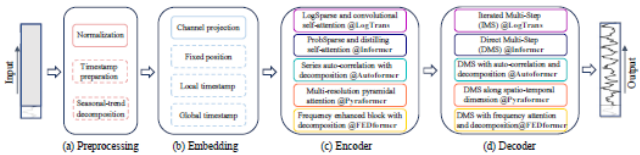

*Figure A1. Transformer-based solutions. Source: Zeng et al. (2022)*

**Figure A2.** Figure A2 presents snapshots of the primary underlying data series from both a randomly selected time window and the full-time span for the two additional datasets. These figures are included to provide insights into the characteristics of the data prior to modeling—such as distribution, stationarity, and trends. A preliminary statistical data analysis was conducted in Google Colab to support this exploration. While this analysis is not part of the core research, it serves as an initial step in understanding the datasets

*Figure A2 – primary time series data for selected time-window.*

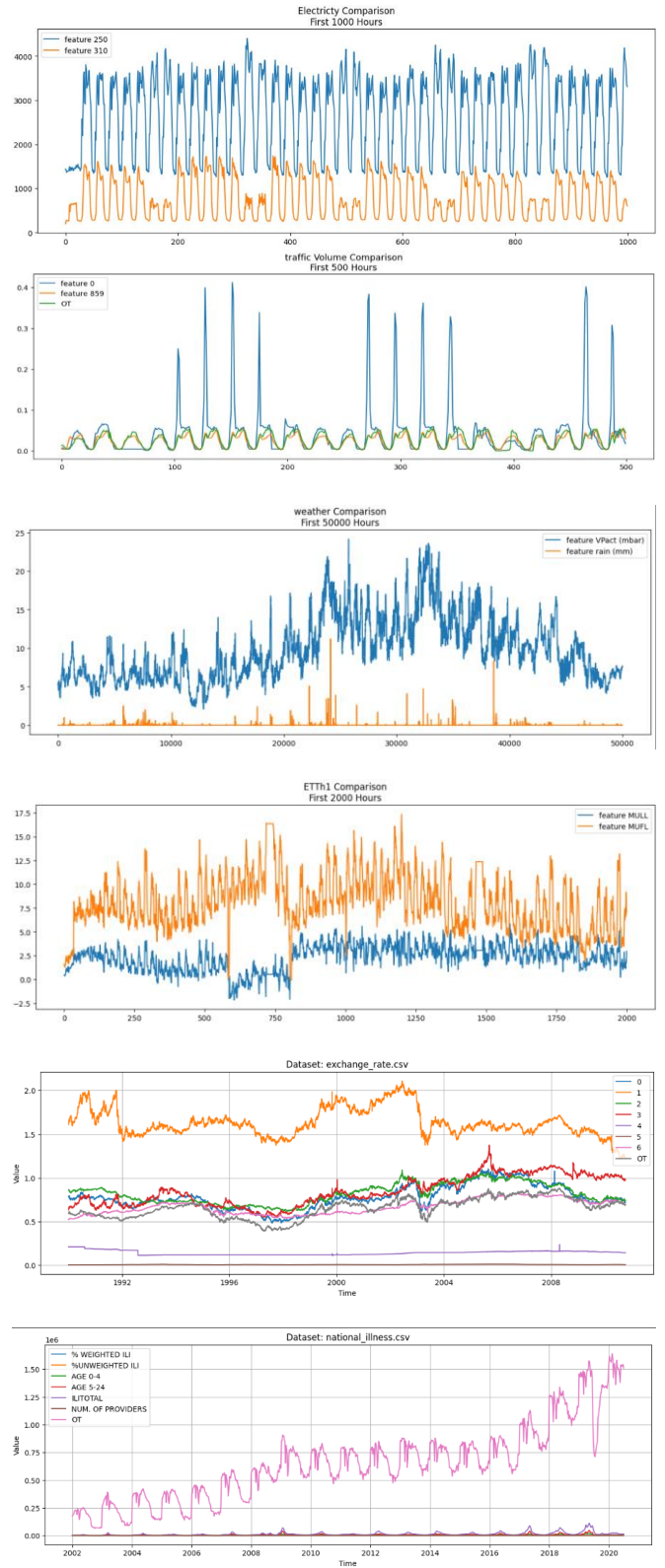

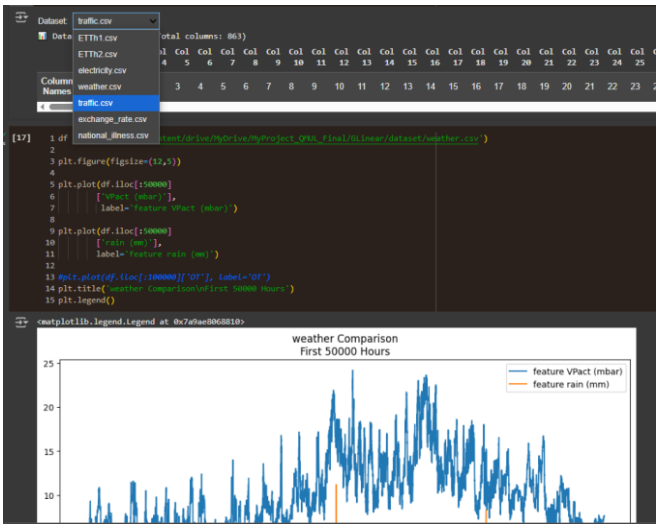